\documentclass{article}

\usepackage{graphicx}      
\usepackage{comment}
\usepackage{float}
\usepackage{subcaption}
\usepackage{color}
\usepackage{siunitx}
\usepackage{amsmath,amssymb}
\usepackage{booktabs}
\usepackage{xcolor}
\newtheorem{ass}{Assumption}

\title{\LARGE \bf
	Constrained reaction wheel desaturation and attitude control of spacecraft with four reaction wheels \footnote{This research is supported by Air Force Office of Scientific Research Grant number FA9550-20-1-0385 and NSF award CMMI-1904394.}
}

\author{Miguel Castroviejo-Fernandez$\;^\dagger$ and Ilya Kolmanovsky
\thanks{University of Michigan, Ann Arbor, MI 48109 USA.
        {\tt\small mcastrov, ilya@umich.edu}. }
}

\begin{document}

\maketitle
\thispagestyle{empty}
\pagestyle{empty}

\begin{abstract}   
The paper addresses a problem of constrained spacecraft attitude stabilization with simultaneous reaction wheel (RW) desaturation. The spacecraft has a reaction wheel array (RWA) consisting of four RWs in a pyramidal configuration. The developments exploit a spacecraft dynamics model with gravity gradient torques. The linearized dynamics are shown to be controllable at almost all RWA configurations. Configurations that result in the highest Degree of Controllability are elucidated. A strategy that combines an incremental reference governor and time-distributed model predictive control is proposed to perform constrained RW desaturation at low computational cost. Simulation results of successful RW desaturation maneuvers subject to spacecraft pointing constraints, RW zero-speed crossing avoidance and limits on control moments are reported.
\end{abstract}

\section{Introduction}

Many spacecraft use reaction wheels (RWs) to maintain pointing and for reorientation (\cite{leve2015spacecraft}). In order to maintain pointing, the reaction wheels absorb externally induced spacecraft total angular momentum changes through rotational speed variations. Thus, they can spin up and eventually need to be desaturated (\cite{bryson1993control}). Traditionally, spacecraft RW desaturation is performed using thrusters to produce moments that result in the decrease of RW rotational speed.  Unfortunately, thrusters consume fuel, a limited resource that constrains spacecraft operational life.

Alternative approaches (see e.g., \cite{bryson1993control}) to RW desaturation include exploiting gravity gradients: moments due to nonuniform gravity force distribution along the body of the spacecraft. With this approach, the desaturation can be performed without the use of thrusters and with zero fuel consumption. This is the approach considered in this paper. 

Additionally, in space missions, constraints are likely to be present: control moments that can be applied are limited, exclusion zones for spacecraft pointing may exist or zero-speed crossing avoidance for RWs could be desirable. Given the presence of constraints, the use of Model Predictive Control (MPC), cf. \cite{rawlings2017model}, is appealing.
In \cite{7330731}, an MPC law is exploited for three RWs desaturation using either gravity gradients or magnetic torques while maintaining spacecraft attitude deviation within a prescribed range.

 Classical MPC formulations require to solve an optimization problem  at each time instant. This can be problematic for spacecraft control applications where the vehicle has limited onboard computing capabilities. Such limitations may be due to the use of slow radiation-hardened processors, to the small size of the spacecraft (CubeSat or SmallSat) or to the need to reduce electrical power consumption.
 The use of suboptimal MPC strategies, such as Time Distributed MPC (TDMPC) \cite{liao2020time} and real-time iterations \cite{diehl2005nominal}, can help reduce the computational burden. In these approaches, a suboptimal solution to the MPC optimization problem is generated through the use, at each time step, of a few iterations of an optimizer and warm-starting.
In \cite{CCTA22}, a spacecraft subject to gravity gradients and equipped with three RWs is considered. An input constrained TDMPC law is used to stabilize spacecraft attitude and concurrently desaturate the RWs. State constraints are not considered beyond providing a conservative estimate of the set of initial conditions for which pointing constraints are not violated.

Even though an RWA with three RWs in orthogonal configuration is sufficient to control spacecraft attitude, in practice, it is common to have the spacecraft equipped with four RWs, as this provides redundancy of the actuators and can help avoid zero-speed crossing of the RWs, see, e.g., \cite{kron2014four}. To the best of our knowledge, controllability of the combined reaction wheels and spacecraft attitude attitude dynamics for a spacecraft subject to gravity gradients and equipped with four RWs has not been studied in the literature. It is unknown if attitude stabilization with concurrent desaturation, while enforcing spacecraft pointing constraints, is possible for such a system.

This paper extends the results in \cite{CCTA22} by addressing the practically relevant case of a spacecraft with four RWs and by adopting an extension of the TDMPC law to handle state constraints on spacecraft pointing.
We demonstrate linear controllability of the combined spacecraft attitude and four RWs dynamics for almost all pyramidal configurations of the RWA and perform a parametric study on the Degree of Controllability of the system. Simulation results demonstrate successful RW desaturation, good closed-loop performance and state constraints handling capabilities.

In this work, the TDMPC is extended using the approach of \cite{ACC23} whereby we augment the input constrained TDMPC  with a variant of a reference governor, see \cite{garone2017reference}. The resulting control scheme is referred to as RG-TDMPC. Starting from a feasible reference command, at each time step, the RG-TDMPC verifies constraint admissibility of an increment in the current reference command and is formulated in such a way that it avoids the need to solve a discrete-time optimal control problem at each time instant over the prediction horizon.  As demonstrated in \cite{ACC23}, the combination of IRG and input constrained Linear Quadratic MPC (LQ-MPC), referred to as RG-MPC,
has the potential to lower computation time as compared to state and input constrained LQ-MPC and has the capability to enforce nonlinear state and control constraints.

The paper is organized as follows.
In Section~\ref{sec:EOM} the spacecraft three-dimensional (yaw-pitch-roll) attitude dynamics model is introduced. Section~\ref{sec:linCtrl} studies the linear controllability properties of the system for different configurations of the RWA. Section \ref{sec:ctrlDesign} describes the implementation of RG-TDMPC.  
Finally, Section~\ref{sec:sim} reports simulation results.

\textbf{Notations:}
Let $\mathbb{S}^n_{++}$, $\mathbb{S}^n_{+}$ denote the set of  symmetric $n\times n$ positive definite and positive  semidefinite matrices respectively. $I_m$ denotes the $m\times m$ identity matrix.
Given $x\in \mathbb{R}^n$ and $W\in \mathbb{S}^n_+$, the W-norm of $x$ is $||x||_W = \sqrt{x^{\top}Wx}$.
Given $a\in\mathbb R^n,\;b\in\mathbb R^m$, $(a,b) = [a^\top,b^\top]^\top$. 
Let $c_{\alpha}= \cos(\alpha)$, $s_{\beta} = \sin(\beta)$. The operator $\Pi_{\mathcal U}(\cdot)$ denotes the projection onto the set $\mathcal U$. A frame $\mathcal A$ is defined by the three orthogonal normalized vectors $x_{\mathcal A}$, $y_{\mathcal A}$, $z_{\mathcal A}$. The cross product of two vectors is denoted by $\times$. The representation of vector $v$ in frame $\mathcal A$ is denoted by $v|_{\mathcal A}$. The sets $\mathbb Z$ and $\mathbb{N}$ are the set of integers and nonnegative integers.

\section{Problem setting}
\label{sec:EOM}
We consider a reaction wheel actuated spacecraft in circular orbit (radius $r_0$) around a celestial body (with gravitational constant $\mu$) and we account for gravity gradients in modeling its dynamics. Gravity gradient induced torques, or simply gravity gradients, are external torques that appear on an object in a gravitational field due to the gravitational force decreasing with the square of the distance. More specifically, a spacecraft orbiting around a celestial body will have a weaker pull on the parts it has further away from the body. As the gravity gradient generated torque is external, it can change the total angular momentum of the spacecraft and hence can, potentially, be used to reduce the RWs rotational speed, i.e. to desaturate them.

\subsection{Model}\label{sec:EOM_values}
\begin{figure}[ht]
\centering
\includegraphics[trim={0 0 0 0}, clip,scale=0.65]{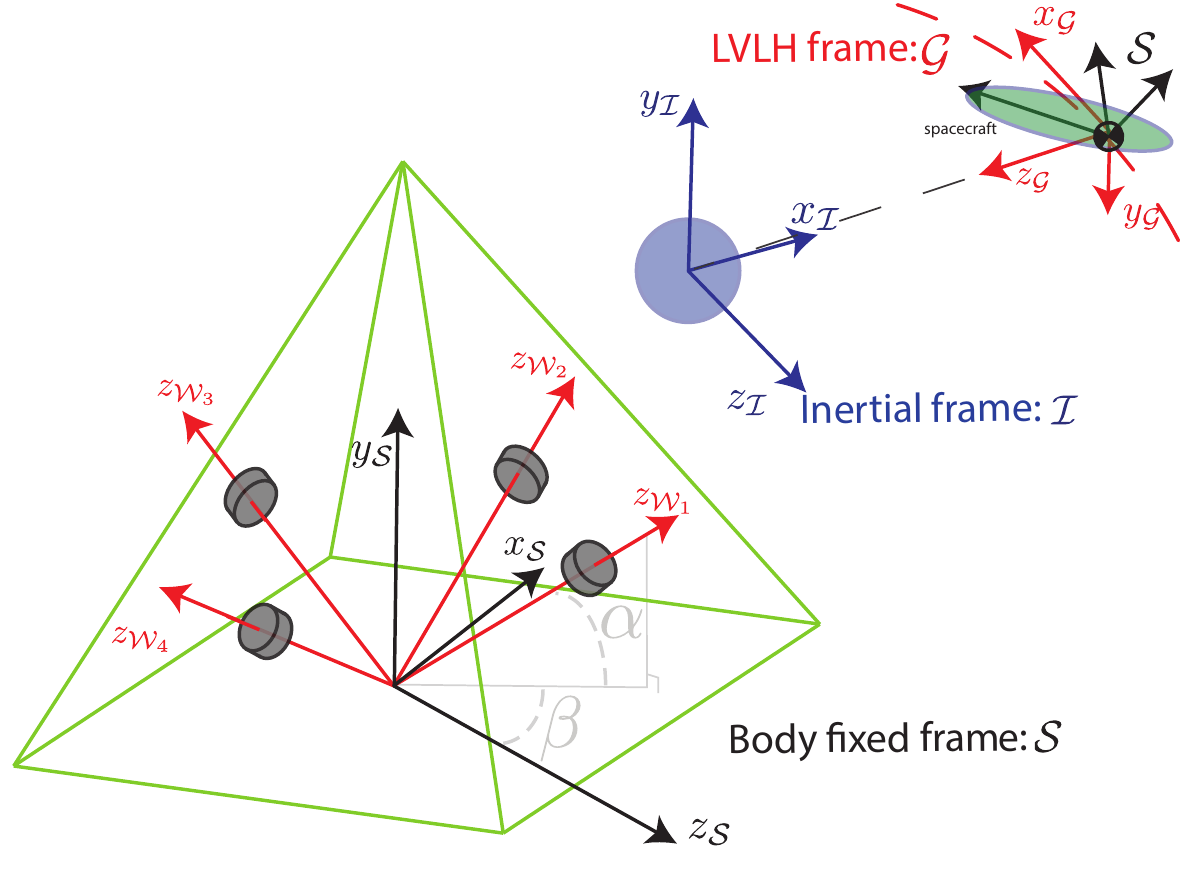}
\caption{Left: pyramidal RWA parameterized by angles $\alpha,\;\beta$.
Right: frames $\mathcal I,\;\mathcal G$ and $\mathcal S$.}
\label{fig:4RW_genConf}
\end{figure}

Let  $\mathcal{I}$ be an inertial frame, $\mathcal{S}$ be a body fixed frame aligned with the principal axes of the spacecraft and $\mathcal{G}$ be a Local Vertical Local Horizontal (LVLH) frame as described in \cite{wie1998space} and depicted in Figure \ref{fig:4RW_genConf} (right). 
The orientation of the body fixed frame $\mathcal{S}$ with respect to LVLH frame $\mathcal{G}$ is specified by  3-2-1 Euler yaw-pitch-roll angles, $\psi,~\theta,~\phi$. The angular velocity of frame $\mathcal S$ with respect to the frame $\mathcal I$ is given by  $\left. \omega_{\mathcal{S}/\mathcal{I}}\right|_{\mathcal{S}} = \begin{bmatrix} \omega_1 & \omega_2 & \omega_3 \end{bmatrix}^{\top}$.
Then, following \cite[Section 6.10]{wie1998space}:
\begin{align}
    \begin{bmatrix}
    \dot \phi\\
    \dot \theta\\
    \dot \psi
    \end{bmatrix} = &\frac{1}{c_\theta }\begin{bmatrix}
    c_\theta  \!&&\! s_\phi s_\theta  \!&&\! c_\phi s_\theta \\
    0 \!&&\! c_\phi c_\theta  \!&&\! -s_\phi c_\theta \\
    0 \!&&\! s_\phi  \!&&\! c_\phi 
    \end{bmatrix}\!\left( \begin{bmatrix}\omega_1\\\omega_2\\ \omega_3\end{bmatrix}\!+\!
     n\!\begin{bmatrix}
    c_\theta s_\psi \\
    s_\phi s_\theta s_\psi  + c_\phi c_\psi \\
    c_\phi s_\theta s_\psi  - s_\phi c_\psi 
    \end{bmatrix}\right)\!, \label{eq:EOM_1}
\end{align}
 where $n = \mu^{1/2} r_0^{-3/2}$ is the circular orbit gravitational parameter. The spacecraft is equipped with an RWA made of four reaction wheels in a typical pyramidal configuration. The RWA is depicted in Figure \ref{fig:4RW_genConf} (left). For each RW, we define a frame $\mathcal W_i,\;i=1,\dots,4$ in which the rotation axis of the corresponding RW is aligned with $z_{\mathcal W_i}$. The frames $\mathcal W_i$ relate to $\mathcal S$ through two parameterized rotations:  
\begin{align*}
 &\mathcal O_{\mathcal W_i/\mathcal S} = O_1(\alpha_i)O_2(\beta_i). \quad i = 1,\dots,4,\\
 &\alpha_i = \alpha,\quad \beta_i = \frac{\pi}{2}(i-1)+\beta,\quad i = 1,\dots,4,
\end{align*}
where $O_j(\cdot),\;j=1,\;2,\;3$ are the cosine matrices around respective axes and $\alpha\in[-\pi/2,\;\pi/2],\;\beta\in[0,\;\pi/2]$ are offset angles characterizing the RWA configuration.
All RWAs considered are symmetric about  $ y_{\mathcal S}$ axis. This is the axis of orbital rotation when $\phi=\theta=\psi = 0$. Furthermore the angular velocity of frame $\mathcal W_i$ with respect to $\mathcal S$ is given by $\omega_{\mathcal W_i/\mathcal S} = \Omega_i z_{\mathcal W_i}$ and we note that $\mathcal O_{\mathcal S/\mathcal W_i}=\mathcal O_{\mathcal W_i/\mathcal S}^\top$.
We make the following assumption.
\begin{ass}
\label{ass:RWAss}
Each RW has its center of mass (c.o.m) along its spin axis. All RWs have identical moment of inertia about the spin axis: $J_s$ and zero transversal moment of inertia.
 The principal moments of inertia of the spacecraft, $J_1,J_2,J_3$, are pairwise distinct.
\end{ass}
The Euler's equations for the rotational dynamics in $\mathcal S$ are 
\begin{equation*}
\label{eq:Euler1}
 J^{sc} \dot \omega_{\mathcal S/\mathcal I}|_{\mathcal S}= 
M^{GG}|_{\mathcal S}\! -\! \omega_{\mathcal S/\mathcal I}|_{\mathcal S} \!\times\! ( J^{sc} \omega_{\mathcal S/\mathcal I}|_{\mathcal S} \!-\!h^{rw}|_{\mathcal S} )\!-\!\dot h^{rw}|_{\mathcal S},
\end{equation*}
where $J^{sc}$ is the inertia matrix of the spacecraft relative to its c.o.m., $\dot\omega_{\mathcal S/\mathcal I}|_{\mathcal S}$ is the time derivative of $\omega_{\mathcal S/\mathcal I}|_{\mathcal S}$ with respect to $\mathcal S$, $M^{GG}|_{\mathcal S}$ are the gravity gradient induced torques, $h^{rw}|_{\mathcal S} = \sum_{i=1}^4 J_{s} \mathcal O_{\mathcal S/\mathcal W_i}\omega_{\mathcal W_i/S}|_{\mathcal W_i}$ is the sum of the angular momenta of each RW relative to its c.o.m. with respect to $\mathcal S$ and $\dot h^{rw} = \sum_{i=1}^4J_{s}\mathcal O_{\mathcal S/\mathcal W_i} \dot\omega_{\mathcal W_i/\mathcal S}|_{\mathcal W_i}$.  After appropriate substitutions, we get
\begin{align}
      & \begin{bmatrix}
    J_1 \dot\omega_1\\J_2 \dot\omega_2\\J_3 \dot\omega_3
    \end{bmatrix} = \begin{bmatrix}
    (J_2 - J_3)(\omega_2\omega_3 - 3n^2C_{2}C_{3}) \\
    (J_3 - J_1)(\omega_1\omega_3 - 3n^2C_{1}C_{3}) \\
    (J_1 - J_2)(\omega_1\omega_2 - 3n^2C_{1}C_{2}) 
    \end{bmatrix}-\omega_{\mathcal{S}/\mathcal{I}}|_{\mathcal{S}}\!
    \times \!\left.h^{rw}\right|_{\mathcal{S}} \nonumber\\
    &+ J_s \begin{bmatrix}
    -c_\alpha s_\beta &-c_\alpha c_\beta &c_\alpha s_\beta &c_\alpha c_\beta\\
    s_\alpha &s_\alpha &s_\alpha &s_\alpha \\
    -c_\alpha c_\beta &c_\alpha s_\beta &c_\alpha c_\beta &-c_\alpha s_\beta
    \end{bmatrix}\begin{bmatrix}
    A_1\\A_2\\A_3\\A_4
    \end{bmatrix}\label{eq:EOM_2},
\end{align}
 where $C_{1} = -s_\theta$, $C_{2} = s_\phi c_\theta$, $C_{3} = c_\phi c_\theta$, $A_{i}$, $i=1,2,3,4$ are the angular accelerations of the RWs and 
 \begin{align*}
          &\left.h^{rw}\right|_\mathcal S =\begin{bmatrix}
          h^{rw}_1\\h^{rw}_2\\h^{rw}_3
          \end{bmatrix} =J_s\begin{bmatrix}
     c_\alpha \left((\Omega_2-\Omega_4) c_\beta+(\Omega_1-\Omega_3)s_\beta\right)\\
     -s_\alpha\left(\Omega_1+\Omega_2+\Omega_3+\Omega_4\right)\\
     c_\alpha\left((\Omega_1-\Omega_3) c_\beta-(\Omega_2-\Omega_4)s_\beta\right)\\
     \end{bmatrix}.
 \end{align*}

The evolution of the RW angular velocities is given by 
\begin{equation}
    \dot \Omega_i = A_i,\quad i = 1,\ldots,4.\label{eq:EOM_3}
\end{equation}

Equations of motion  (\ref{eq:EOM_1})-(\ref{eq:EOM_3}) can be aggregated into  
\begin{subequations}
\label{eq:NLmodelFull}
\begin{align}
&\dot{{x}} = {f}({x},{u}),\label{eq:NLmodelFullx}\\
&y= [0_{2\times 8}\;I_2]x,
\end{align}
\end{subequations}
where ${x} = [\phi,\; \theta,\; \psi,\; \omega_1,\; \omega_2,\; \omega_3,\; \Omega_1,\;\Omega_2,\;\Omega_3,\;\Omega_4]^{\top}$, $  {u} = [A_1,\; A_2,\; A_3,\;A_4]^{\top}$ and the outputs $y$ are chosen so that they can be commanded to track set-points. Define
\begin{equation}
    x^{eq}(v) =
    \begin{bmatrix}
    0&0&0&0&-n&0&a&b&a&b
    \end{bmatrix}^\top,
    \label{eq:n1EQ}
\end{equation}
where $v = [a,b]^\top \in \mathbb R^2$. Plugging in (\ref{eq:n1EQ}) and a zero input into equations (\ref{eq:NLmodelFull}) results in $\dot x = 0$ and $y = v$. Thus, (\ref{eq:n1EQ}) represents a family of unforced equilibria for system (\ref{eq:NLmodelFull}) and all $(\alpha,\;\beta)\in[\frac{-\pi}{2},\;\frac{\pi}{2}]\times[0,\;\frac{\pi}{2}]$.

The set of equilibria (\ref{eq:n1EQ}) characterizes a spacecraft in circular orbit, with axis of orbital rotation along $y_{\mathcal S}$, and oriented so that the effect of the gravity gradients is null. At each of these equilibria, the motion is thus described by the 2 Body Problem without perturbation forces. In this case, the total angular momentum of the spacecraft is conserved and should be along $y_\mathcal S$. It then follows that the angular momentum of the RWA is required to be zero along $x_\mathcal S$ and $z_\mathcal S$ ( $h^{rw}_1= h^{rw}_3 =0$), which translates into the values $x_i^{eq}(v)$, $i=7,8,9,10$.  
The concurrent RWA desaturation and spacecraft attitude dynamics stabilization can be stated as the reference tracking problem of bringing the states of the system (\ref{eq:NLmodelFull}) to an equilibrium $x^{eq}(r)$ associated to $r\in\mathbb R^2$ starting from some initial condition (IC), $x^0$.

\subsection{Numerical values and constraints}
\label{sec:numVals}
In the subsequent analysis and simulations, unless otherwise specified, we consider a spacecraft in circular orbit around Earth ($\mu = 3.986\times 10^{5}\;\tt km^3/ \tt s^{2}$), at an altitude of $500\;\tt km$ ($n = 1.1086 \times10^{-3}\;\tt s^{-1}$) and an orbital period $T_{orb} = \frac{2\pi}{n}\approx1.58\;\tt hr$. The spacecraft principal moments of inertia are $J_1 = 1000,\;J_2 = { 2200}$ and $ J_3 = 1400\;\tt kg\;\tt m^2 $. Each RW moment of inertia about its spin axis is  $J_s = 0.1\;\tt kg\;m^2$. Additionally, we consider the following constraints: 
\begin{itemize}
    \item Input saturation: The electric motors have a maximum torque $\tau_{max} = 0.05\;\tt Nm$ translating into a maximum RW acceleration limits as $|A_i|=|u_i| \leq \tau_{max}/J_s = 0.5\;{\tt rad\;s^{-2}},\;i=1,2,3,4$.
    \item Attitude pointing: Pointing requirements on the spacecraft during reaction wheel desaturation maneuver are given as $|x_i|\leq 0.1\;{\tt rad},\;i=1,2,3$.
    \item Zero-speed crossing: Zero crossing by RW speed must be avoided during desaturation: $sign(x_i)=sign(x^0_i),\;i=7,8,9,10$.
\end{itemize}
Let $\mathcal U,\;\mathcal X$ represent the set of admissible inputs and states, respectively.

\section{Linear Controllability Analysis}
\label{sec:linCtrl}
Linearizing (\ref{eq:NLmodelFullx}) about $x^{eq}(r)$ results in
\begin{equation}
\label{eq:linSys}
    \dot {\tilde x} = A_{r} \tilde x + B_{r}  u , \;\tilde x = x-x^{eq}(r),
\end{equation}
where
\begin{subequations}
\label{eq:ABMatrices}
\begin{align}
\!&\!A_{r}\! = \!E\!\left[\!\begin{array}{*{10}c}
0\!&\!0\!&\!n\!&\!1\!&\!0\!&\!0\!&\! \!&\! \!&\! \!&\! \\
0\!&\!0\!&\!0\!&\!0\!&\!1\!&\!0\!&\! \!&\!  0_{3\times 4}\!&\! \!&\! \\
-n\!&\!0\!&\!0\!&\!0\!&\!0\!&\!1\!&\! \!&\! \!&\! \!&\! \\
a_{4,1}\!&\!0\!&\!0\!&\!0\!&\!0\!&\!a_{4,6}\!&\!  c_{\alpha}c_{\beta}\!&\! -c_{\alpha}s_{\beta}\!&\! -c_{\alpha}c_{\beta}\!&\! c_{\alpha}s_{\beta} \\
0\!&\!a_{5,2}\!&\!0\!&\!0\!&\!0\!&\!0\!&\!0\!&\!0\!&\!0\!&\!0 \\
0\!&\!0\!&\!0\!&\! a_{6,4}\!&\!0\!&\!0\!&\! -c_{\alpha}s_{\beta}\!&\! -c_{\alpha}c_{\beta}\!&\!  c_{\alpha}s_{\beta}\!&\! c_{\alpha}c_{\beta}\\
\!&\! \!&\! \!&\! \!&\! \!&\!  \!&\! \!&\! \!&\! \!&\!\\
\!&\! \!&\! \!&\! \!&\! \!&\! \!&\!0_{4\times 10} \!&\! \!&\! \!&\!\\
\!&\! \!&\! \!&\! \!&\! \!&\!  \!&\! \!&\! \!&\! \!&\!
\end{array}\!\right]\!,\\
  &  B_{r} = \!E
 \!\begin{bmatrix}
0_{3\times 4}\\
\begin{matrix}
 \frac{-c_{\alpha}s_{\beta} }{n}& \frac{- c_{\alpha}c_{\beta} }{n}& \frac{ c_{\alpha}s_{\beta} }{n}&\frac{  c_{\alpha}c_{\beta}}{n}\\
 \frac{s_{\alpha} }{n}&\frac{ s_{\alpha} }{n}&\frac{  s_{\alpha} }{n}& \frac{             s_{\alpha}}{n}\\
\frac{- c_{\alpha}c_{\beta} }{n}& \frac{  c_{\alpha}s_{\beta} }{n}& \frac{ c_{\alpha}c_{\beta} }{n}& \frac{- c_{\alpha}s_{\beta}}{n}
\end{matrix}\\
I_{4\times 4}
 \end{bmatrix}\!,
\end{align}
\end{subequations}
\begin{align*}
&E = diag\left(\begin{bmatrix}
 1_{1\times 3}&\frac{nJ_s}{J_1}&\frac{nJ_s}{J_2 } &\frac{nJ_s}{J_3 }& 1_{1\times 4}
\end{bmatrix}\right)\\
    &a_{4,1 }=-3n(J_2 - J_3)/J_s,\; a_{5,2}=-3n(J_1 - J_3)/J_s,\\
    &a_{4,6} = (J_3 - J_2 )/J_s -2 s_{\alpha} \begin{bmatrix}    1&1    \end{bmatrix}r/n,\\
   &a_{6,4}= (J_2 - J_1 )/J_s + 2 s_{\alpha} \begin{bmatrix} 1&1 \end{bmatrix}r/n.
\end{align*}

Using Matlab, the controllability matrix of system (\ref{eq:linSys}),
defined as $\mathcal C = [B,\;AB,\dots,A^9B]$,
and its rank were computed for  $(\alpha,\;\beta)\in\left\{[-90,\;90]\cap\mathbb Z\right\}\times\left\{[0,\;90)\cap\mathbb Z\right\}$ $[{\tt deg\times deg}]$ while keeping all other parameters appearing in matrices (\ref{eq:ABMatrices}), including $r$, symbolic. When $\alpha\notin\{-90^\circ,\;0^\circ,\;90^\circ\}$, the linearization of system (\ref{eq:NLmodelFull}) at any equilibrium point described by (\ref{eq:n1EQ}) is controllable. Moreover, when $\alpha = 0^\circ$, the rank of the controllability matrix is eight and when $\alpha = \pm \frac{\pi}{2}$ it is equal to six. Setting $\alpha=\pm90^\circ$ in (\ref{eq:ABMatrices}) and reordering the state vector as $ \nu_i = \tilde x_j,\;i=\{1,\dots,10\},\;j=\{2,\;5,\;7,\;8,\;9,\;10,\;1,\;3,\;4,\;6\}$ yields a controllable-uncontrollable realization of the system where the last four states describe the uncontrollable subsystem. As the change of basis to obtain 
$\nu$ from $\tilde x$ is a permutation, the uncontrollable modes of system (\ref{eq:linSys}) can be deduced directly. They represent motions along $\tilde x_1,\;\tilde x_3,\;\tilde x_4$ or $\tilde x_6$. These states are all in the $x_\mathcal S-z_\mathcal S$ plane, which is reasonable given that $\alpha = \pm90^\circ$ corresponds to all RWs along the $y_{\mathcal S}$ axis. Similarly, $\alpha =0^\circ$ implies that all the RWs spin axes lie in the $x_{\mathcal S}-z_{\mathcal S}$ plane. A similar analysis shows that states $\tilde x_2,\;\tilde x_5$ are completely decoupled from the inputs.

\subsection{Estimating the Degree of Controllability}

As a metric for the Degree of Controllability (DoC) we use,
\begin{equation} 
    J_{ind}(t_f,t_0) = \lambda_{max}\left(e^{A^\top(t_f-t_0)}M(t_f,t_0)^{-1}e^{A(t_f-t_0)}\right), \label{eq:JindEq}
\end{equation}
where $M(t_f,t_0)$ is the controllability gramian over the finite time interval $[t_0,\;t_f]$ and $\lambda_{max}(\cdot)$ denotes the maximum eigenvalue. The metric $J_{ind}$ characterizes, in a $2$-norm squared sense, the maximum control effort needed
 to bring an initial state of unit norm to the origin in $\Delta T = t_f-t_0$ sec.  The normalized eigenvector,  $\tilde x_{ind}$, associated to the maximum eigenvalue in (\ref{eq:JindEq}) corresponds to the IC, among those of unit norm, that requires the maximum effort to bring the state of the system to the origin starting from it.

\begin{figure}[ht]
\centering
\includegraphics[scale=0.65]{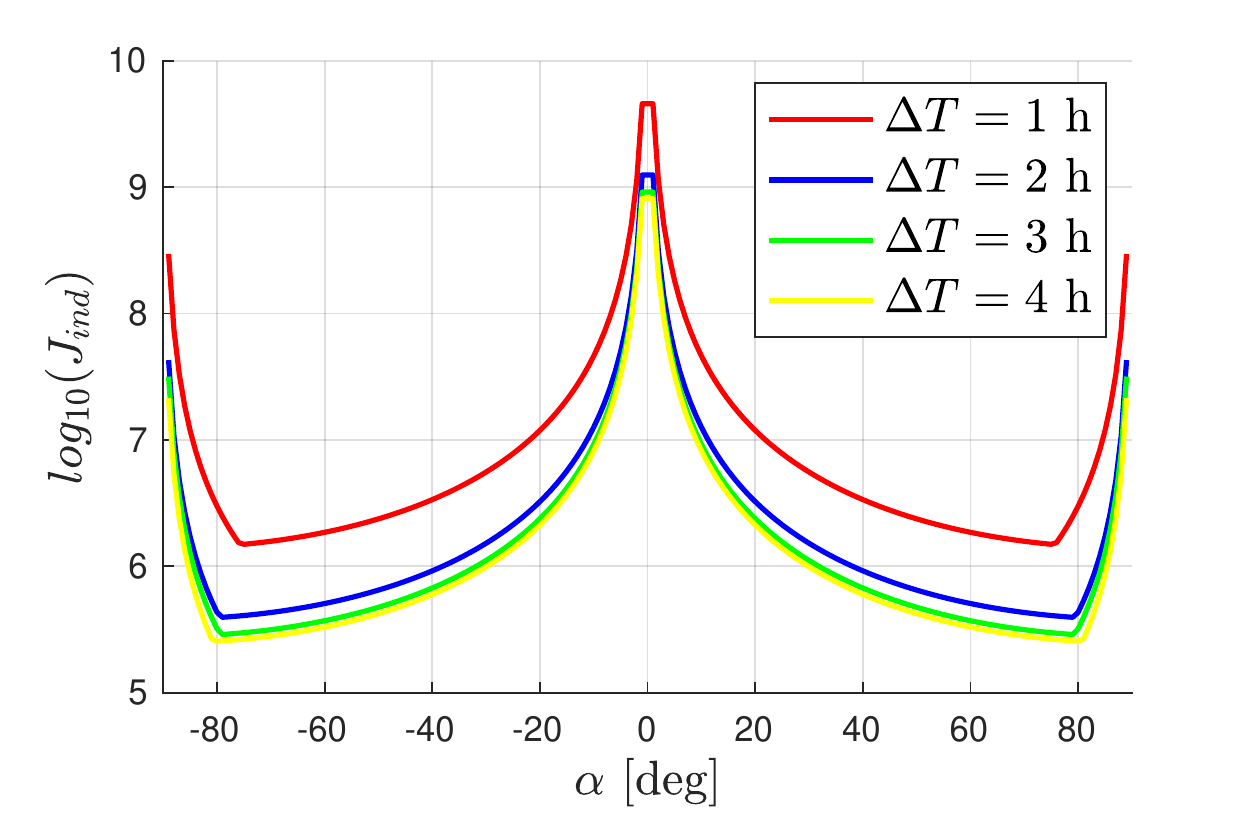}
\caption{$log_{10}$ of the DoC for different values of $\alpha$ and $\Delta T$. The DoC was not computed at the singular values $\alpha=0,\;\pm 90^o$.}
\label{fig:DoC_ths}
\end{figure}

For the spacecraft parameter values in Section \ref{sec:numVals},
$J_{ind}(t_f,t_0)$ is symmetric about $\alpha = 0$ and is a weak function of  $\beta$.
Figure \ref{fig:DoC_ths} shows $log_{10}(J_{ind})$ for different values of $\Delta T$, over a range $\alpha\in(-90^o,\;90^o)$ and $\beta= 0$. 
As $\alpha$ tends to the values  $\{-90^o,\;0^o,\;90^o\}$ the required control effort increases.
Furthermore, increasing the maneuver time reduces the control effort required. This dependence decreases for large maneuver time.  In particular, almost no difference can be seen between the three and four hours maneuver. The values of $\alpha$ and $J_{ind}$ corresponding to the minimum of $J_{ind}$  are denoted by $\alpha_{\tt min}$ and $J_{ind,\tt min}$ respectively. When $\Delta T$ increases from 1 to 4 hours, $\alpha_{\tt min}$ shifts from $76^o$ to $80^o$. The high values of $\alpha_{\tt min}$ can be related to the largest moment of inertia of the spacecraft being along the $y_{\mathcal S}$ axis and to an increased influence of the inputs on the $y_{\mathcal S}$ related states as $\alpha$ approaches $90^o$. By observing $\tilde x_{ind}$, we see that the IC that requires the most effort is 
\begin{align*}
        &\tilde x_{ind} \approx[0_{1\times 4}, 1 ,0_{1 \times 5}]^\top\text{, if } \alpha< \alpha_{\tt min},\\
        &\tilde x_{ind} \approx[0_{1\times 3}, 0.66, 0 , 0.75,0_{1 \times 4}]^\top \text{, if } \alpha\geq \alpha_{\tt min}.
\end{align*}
 For values below $\alpha_{\tt min}$, $\tilde x_{ind}$ is directed purely along $\omega_2$. Above $\alpha_{\tt min}$, $\tilde x_{ind}$ has non zero components of $\omega_1,\;\omega_3$. In both cases, however, the values of RW angular velocities are close to zero, this is due to the difference between $J_s$ and $J_1,\;J_2,\;J_3$ (several orders of magnitude). To understand if a specific combination of RW speeds is harder to desaturate than others one can compute $J_{ind}$ for the subsystem corresponding to states $\tilde x_{i}, i = 7,8,9,10$. This is equivalent to considering ICs that have the first six states at the origin in the computation of the effort. In that case and for all tested  values of $\alpha$ and $\beta$, the highest effort occurs when $\Omega_{i} =0.5$, $i=1,2,3,4$.

 To study how the values of the moment of inertia influence the DoC we set $J_3 = 1150 \tt kg\;m^2$ and vary either $J_1$ or $J_2$ in the range $[300,\;2200]\;\tt kg\;m^2$ while keeping the remaining moment of inertia at $1050 \tt kg \;m^2$. We compute the effort for different values of $\alpha$ and look for $\alpha_{\tt min}$ assuming a maneuver time ($\Delta T$) of one hour. The values $1050,\;1150\;\tt kg\;\tt m^2$ are selected so that the moments of inertia that are kept constant are close but do not entail any controllability issues. Also, selecting values that are close enables us to vary the remaining moment of inertia over a large range without violating the triangle inequality for the moments of inertia.
 
\begin{figure}[h] 
  \begin{minipage}[b]{0.5\linewidth}
    \centering
    \includegraphics[width=1\linewidth]{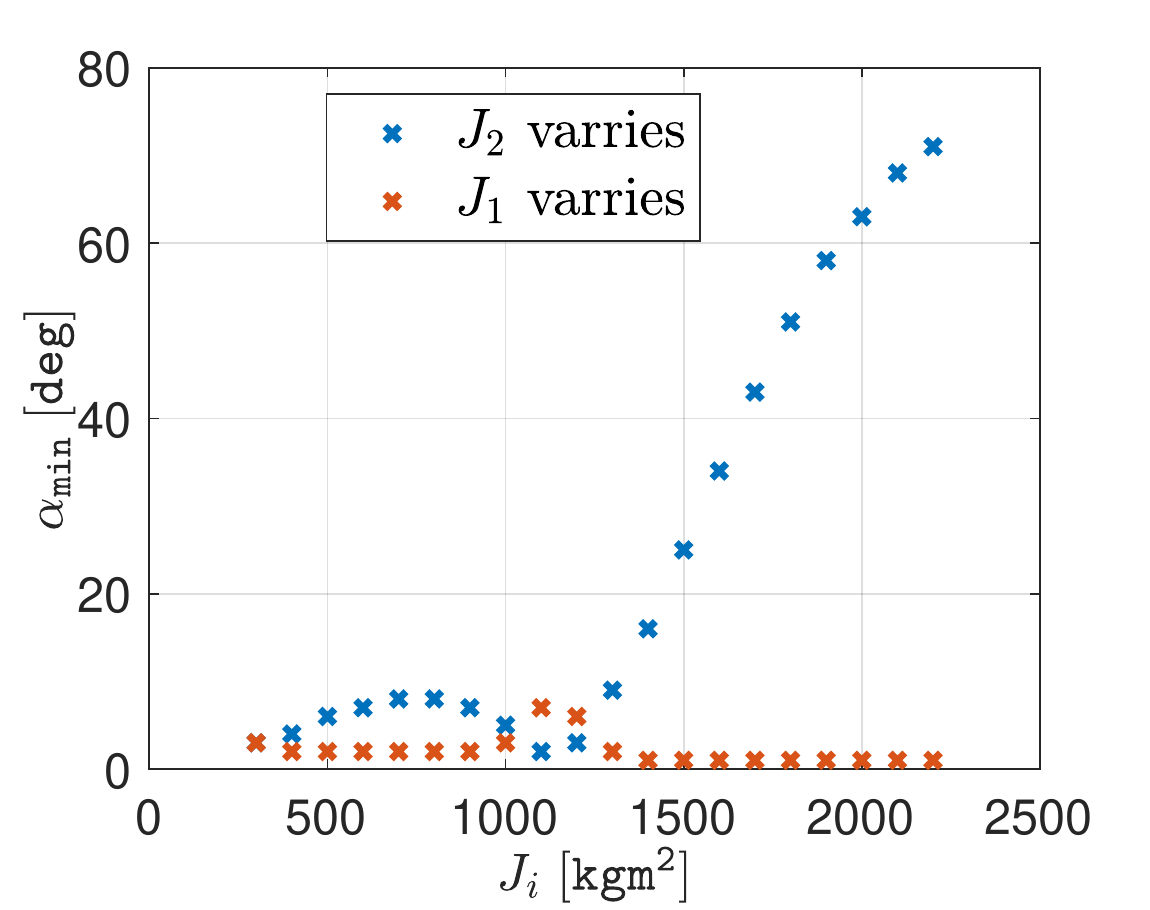} 
  \end{minipage}
  \begin{minipage}[b]{0.5\linewidth}
    \centering
    \includegraphics[width=1\linewidth]{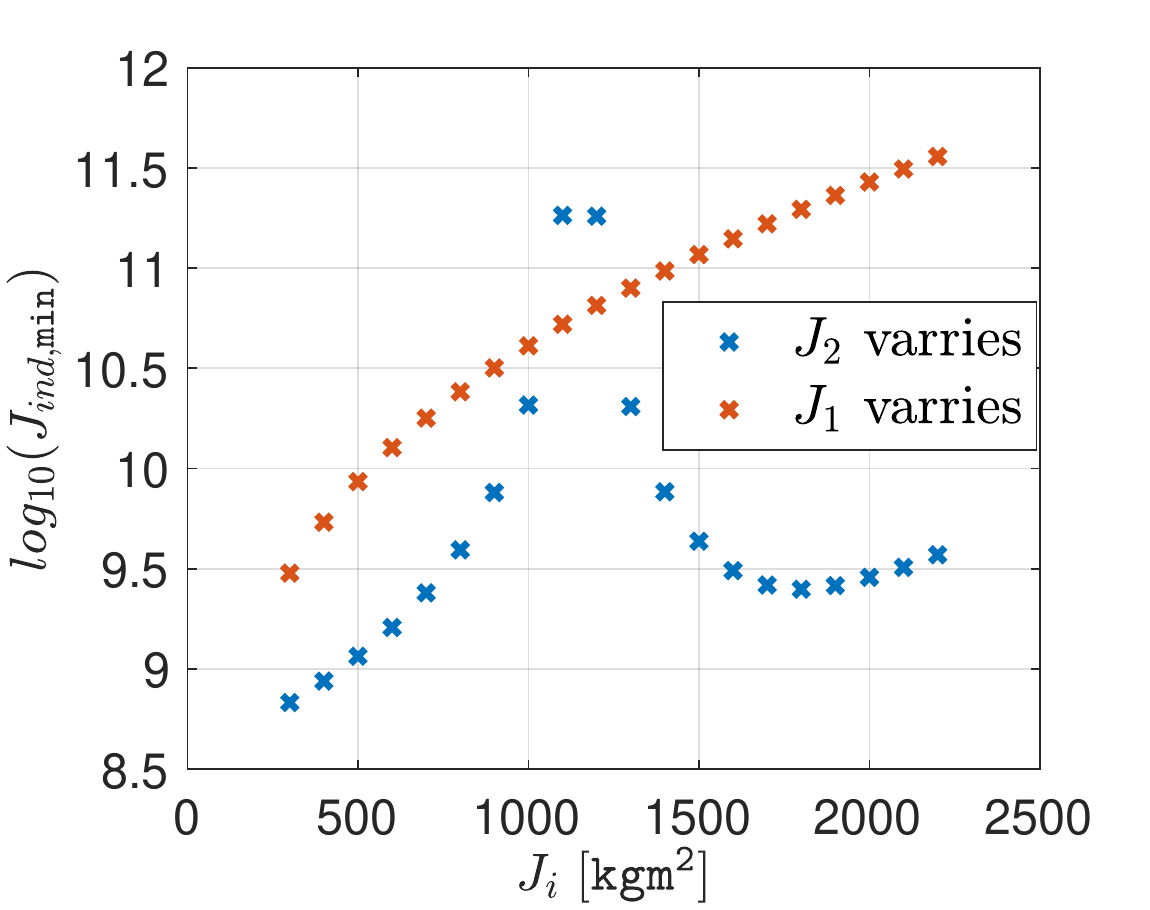} 
  \end{minipage} 
  \caption{$\alpha_{\tt min}$(left) and $log_{10}(J_{ind,\tt min})$(right) as a function of the spacecraft principal moments of inertia.}
  \label{fig:DoC_alphaMin}
\end{figure}

 Figure \ref{fig:DoC_alphaMin} shows the evolution of $\alpha_{\tt min}$ and $\log_{10}(J_{ind,\tt min})$ for different values of the principal moments of inertia. One can observe that $\alpha_{\tt min}$ is strongly dependent on the largest moment of inertia. More precisely, when varying $J_1$, the axis corresponding to the largest moment of inertia is always in the $x_{\mathcal S}$-$ z_{\mathcal S}$ plane and the angle $\alpha_{\tt min}$ is always close to zero. Varying $J_2$ leads to an increase of $\alpha_{\tt min}$ once $J_2$ becomes dominant over $J_1$, $J_3$ ($J_2\geq 1200$). When looking for the maximum control effort (Figure \ref{fig:DoC_alphaMin} right) we observe a vertical asymptote appear when $J_2$ is close to the value of $J_3 = 1150$.  This is in agreement with a loss of linear controllability when either $J_1$ or $J_2$ are equal to $J_3$ (tested for the set of ($\alpha,\beta$) presented at the start of Section \ref{sec:linCtrl}). No loss of controllability was observed when $J_2 = J_1$. 

\section{RG-TDMPC}
\label{sec:ctrlDesign}

For the concurrent RWA desaturation and spacecraft attitude dynamics stabilization subject to state constraints we exploit the RG-MPC scheme introduced by \cite{ACC23} for linear discrete-time systems.  This scheme combines input-constrained LQ-MPC for tracking the applied reference command, $v$, and an incremental reference governor to govern $v$ towards the desired reference command $r$ in a way that satisfies state constraints. As input-constrained LQ-MPC can be made computationally fast and anytime feasible, RG-MPC has been shown to lower the overall computational effort as compared to state and input constrained LQ-MPC
while being capable of enforcing nonlinear state and input constraints. 

The implementation of RG-MPC in this paper is based on the linearized discrete-time model of spacecraft dynamics  given by
\begin{equation}
     \bar{x}_{k+1} = \bar A_r  \bar{x}_k + \bar B_{r} u_k,\; 
     \label{eq:DTLTIsys}
\end{equation}
where $\bar{x}_k = x_k - x^{eq}(v_k),$ $v_k$ is the apllied reference command, the sampling period is $T_s$, and discrete time instants are $t_k=kT_s$.  To simplify the implementation, 
the matrices $\bar{A}_r$, $\bar{B}_r$ correspond
to the linearization of (\ref{eq:NLmodelFullx}) 
at $x^{eq}(r)$ rather
than $x^{eq}(v_k)$ as the differences between
linearizations are small and our final simulations
are based on the nonlinear model. 

An input-constrained LQ-MPC problem for (\ref{eq:DTLTIsys}) is defined as
\begin{subequations}
        \label{eq:IRG_MPC1}
    \begin{alignat}{2}
    \min_{\{\xi_j\},\{\mu_j\}}&~ &&\sum_{j = 0}^{N_{\tt MPC}-1} \|\xi_j \|_Q^2 + \|\mu_j\|_R^2  +\|\xi_{N_{\tt MPC}}\|_{P_r}^2 \\
    \text{s.t.}&~ &&~ \xi_0 = x_k- x^{eq}(v_k)\\
    & &&~\xi_{j+1} = \bar A_{r}\xi_j + \bar B_{r}\mu_j,\; j = 0,\dots,N_{\tt MPC}-1,\\
    & &&~\mu_j \in \mathcal{U},\quad j = 0,\dots,N_{\tt MPC}-1.
    \end{alignat}
    \end{subequations}
where, $\mathcal U$ is the set representing input constraints, $Q\in\mathbb{S}^n_{++},$ $R\in\mathbb{S}^m_{++},$ $P_r \in\mathbb{S}^n_{++}$ is the solution to the Discrete Algebraic Riccati Equation $P_r=Q+\bar A_r^\top P_r \bar A_r-\bar A_r^\top P_r \bar B_rK_r$, and
\begin{equation}
K_r=(R+\bar B_r^\top P_r\bar B_r)^{-1}(\bar B_r^\top P_r\bar A_r),\label{eq:LQRGain}\end{equation}
is the associated LQR gain. Solving (\ref{eq:IRG_MPC1}) generates an optimal control sequence, 
$$\{u_j^{\tt MPC}(x_k,v_k)\}_{ j=0}^{N_{\tt MPC}-1}.$$
The RG-MPC relies on an extension of the MPC generated input sequence given by 
\begin{equation}
 u_j^{\tt ext}(x_k,v_k) =\begin{cases}
     u^{\tt MPC}_j(x_k,v_k)       & \text{if } j < N_{\tt MPC},\\
    \Pi_{\mathcal U}\left(K_r \bar{x}^{ext}_j\right) &  \text{if } j  \geq N_{\tt MPC},
  \end{cases}\label{eq:IRG_MPCSEC1}
\end{equation} 
where,  $K_r$ is the LQR gain in (\ref{eq:LQRGain}) and 
\begin{align*}
    &\bar{x}^{\tt ext}_0(x_k,v_k) = x_k - x^{eq}(v_k),\\
    &\bar{x}^{\tt ext}_j(x_k,v_k) = \bar A_r^j \bar{x}^{\tt ext}_0 + \sum_{i=0}^{j-1}\bar A_r^{j-1-i}\bar B_r u^{\tt ext}_i, j\geq 1,
\end{align*}
 is the associated state sequence. 

Let the reference command at a time instant $t_{k-1}$ be $v_{k-1}$. 
At the next time instant, $t_{k}$, the RG-MPC computes an incremented reference command $v^+$ and checks the constraint admissibility of the state sequence $\bar{x}_j^{\tt ext}(x_k,v^+)$ over a prediction horizon $N_{\tt RG}\geq N_{\tt MPC}$.  Specifically, the following conditions are checked:
\begin{subequations}
\label{eq:IRG_refCond2}
\begin{align}
 & \bar{x}^{\tt ext}_j(x_k,v^+) + x^{eq}(v^+) \in \mathcal{X}, \; j=0,\dots, N_{\tt RG}-1, \\
 &\bar{x}^{\tt ext}_{N_{\tt RG}}(x_k, v^+) \in \mathcal{I}^{\tt LQR}(v^+).\label{eq:IRG_refCondSub2}
 \end{align}
\end{subequations}
If conditions (\ref{eq:IRG_refCond2}) are satisfied for the current state and $v^+$ then $v_k =v^+$, otherwise $v_k = v_{k-1}$. In (\ref{eq:IRG_refCondSub2}), the set $\mathcal{I}^{\tt LQR}({v^+}) \subset \mathbb R^n$ is defined for the reference command $v^+$ and system (\ref{eq:DTLTIsys}) controlled by the LQR law (\ref{eq:LQRGain}). It is  forward invariant for the dynamics $\xi_{k+1} = (\bar A_r+\bar B_r K_r)\xi_k$ and constraint admissible, i.e.,
$\xi+x^{eq}(v^+) \in \mathcal{X},\;K_r x\in\mathcal U$ for all $\xi \in \mathcal{I}^{\tt LQR}(v^+)$. Note that the constraint verification procedure can be terminated earlier if the predicted trajectory enters $\mathcal I^{\tt LQR}(v^+)$ before $N_{\tt RG}$ steps.

The incremented reference command, $v^+$, is formed using the initial, preceding and desired reference commands: $v_0$, $v_{k-1}$ and $r$, respectively. Specifically,
\begin{equation}
    v^+ = \Pi_{\mathcal{L}} \left[ v_{k-1} +  \Delta(v_0,r) \right],
\end{equation} 
where
$\Delta(v_0,r)$ is the reference increment (chosen in our case as described in Section~\ref{sec:sim}),
$$\mathcal{L}={\tt convh}\{v_{0,1},r_{1}\} \times \cdots \times
{\tt convh}\{v_{0,n_v},r_{n_v}\},
$$
$n_v$ is the dimension of $v$ and ${\tt convh}$ denotes the convex hull.

Assuming that $v_0$ is admissible at $t_0$, the control action at time instant $t_k$ is determined as 
$$
u_k=\left\{\begin{array}{cc}
              u^{\tt ext}_{k-k'}(x_{k'},v_{k'}) & \mbox{if $k-k'<N_{\tt MPC}$,} \\
              \Pi_{\mathcal U} \left[K_r(x_k - x^{eq}(v_k)) \right]
              & \mbox{otherwise},
              \end{array}
              \right.
$$
 where $0\leq t_{k'}\leq t_k$, is the last time instant at which $v^+$ was found to be admissible.   

In this paper, we compute sequence (\ref{eq:IRG_MPCSEC1}) using the suboptimal solution  produced by the TDMPC (\cite{liao2020time}) which uses a specified number of iterations, $l$, of primal projected gradient and warm starting.  For this reason, we refer to this control scheme as RG-TDMPC.

Note that the extended control sequence satisfies the control constraints by construction. In particular, box constraints on the inputs are straightforward to satisfy with simple saturation. Also, the predicted state trajectory, $\bar x^{\tt ext}(x,v)$, can be predicted more accurately by forward propagation of the nonlinear model instead of the linearized model. However, this may entail higher computational effort. 

\section{Simulated RW Desaturation Maneuvers with RG-TDMPC}
\label{sec:sim}
An empirical estimate of the necessary number of iterations to maintain closed-loop asymptotic stability of TDMPC with primal projected gradient solver, $l_{\tt min}$, was determined following a similar procedure to that described in \cite[section IV, A]{CCTA22}. For the numerical values presented in Section \ref{sec:numVals} and $N_{\tt MPC} = 5$, $T_s=10$ sec, $Q=diag([1_{1\times6},\;10^{-4}\times 1_{1\times 4}]),$ and $ R = 10^{-8}\times I_4$, the estimate was found to be $l_{\tt min} = 6$. TDMPC performs $l = l_{min}$ iterations of the optimizer per time step unless otherwise specified. For RG-TDMPC, we consider a reference increment,
$\Delta(v_0,r)=\left[\begin{array}{cc} \Delta_1 & \Delta_2 \end{array} \right]^\top$,
defined by 
 $$ \Delta_j =  \frac{0.3(v_{0,j}-r_j) |v_{0,j}-r_j|}{
   (\max_{i\in\{1,2\}}\{|v_{0,i}-r_i|\})^2},~j= 1,2,$$
where $r$ is the final reference and $v_0$ is chosen as the average of initial RW speeds in each RW pair, 
$$v_0 = (\frac{x_{0,7}+x_{0,9}}{2},\frac{x_{0,8}+
x_{0,10}}{2}).$$ 
The terminal constraint set in (\ref{eq:IRG_refCondSub2}) is defined as 
\begin{equation}
    \mathcal I^{\tt LQR}(v) = \{x\!\in\!\mathbb R^{10}|(x\!-\!x^{eq}(v))^\top P_F (x\!-\!x^{eq}(v))\leq c_F\},\! \label{eq:TSetSim}
\end{equation}
where $P_F$ is the solution to the Lyapunov equation $(\bar A_r+\bar B_r K_r)^\top P_F(\bar A_r+\bar B_r K_r) - P_F + I_{10} = 0$ associated with system (\ref{eq:DTLTIsys}) under the LQR feedback law (\ref{eq:LQRGain}). The value $c_F = 10^{10}$ is chosen so that all LQR controlled trajectories starting from an element of the set do not reach RW speeds deviating by more than 1 $\tt rad/s$ from the reference and do not have input or pointing constraints violations. When $v_k\neq r$, we set $N_{\tt RG}=3000$. For possible reduction in computation time, while evaluating (\ref{eq:IRG_refCond2}) we check for inclusion of the predicted trajectory in $\mathcal I^{\tt LQR}(v^+)$ after $N_{\tt MPC}$ steps and every $50$ steps thereafter. When $v_k=r$,  we set $N_{\tt RG} = N_{\tt MPC}$.

\subsection{Comparison to TDMPC}
\begin{figure}[ht]
\centering
\includegraphics[trim={0 0 0 0}, clip,scale=0.7]{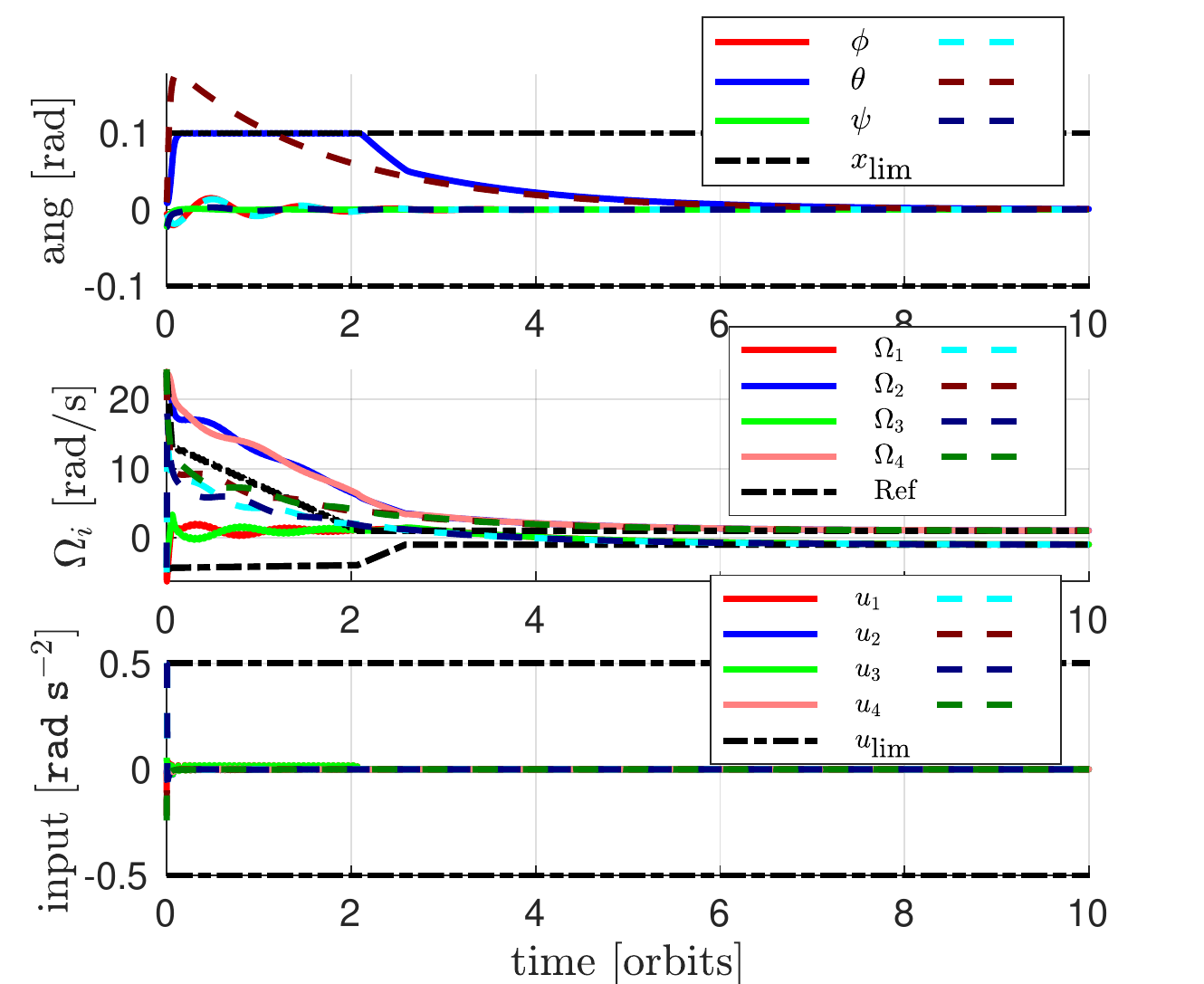}
\caption{State and input trajectories during the RW desaturation maneuver starting from $x_0^\top = [-0.006,0.009,-0.023,0-n,0,-5,23.5,-4.4,24.3]$ to the final reference $x^{eq}([-1\;1]^\top)$ using TDMPC (dashed lines) or RG-TDMPC  with pointing constraints (solid lines). In both cases, $l=6$.}
\label{fig:4RWDesat}
\end{figure}

To begin, we compare RG-TDMPC to TDMPC. The latter was the controller used in \cite{CCTA22} for a similar system.
Figure \ref{fig:4RWDesat} shows the state and input histories during a desaturation maneuver to $x^{eq}([-1,\;1]^\top)$ using either TDMPC (dashed lines) or RG-TDMPC accounting for pointing constraints (solid lines). The desaturation is achieved primarily by pitching the spacecraft. Although the input constraints are enforced, $x_2$ reaches values higher than $0.1\;\tt rad$ with TDMPC. With RG-TDMPC, pointing constraints are enforced. The time required for desaturation is comparable in both cases.

\begin{figure}[ht]
\centering
\includegraphics[trim={0 0 0 0}, clip,scale=0.7]{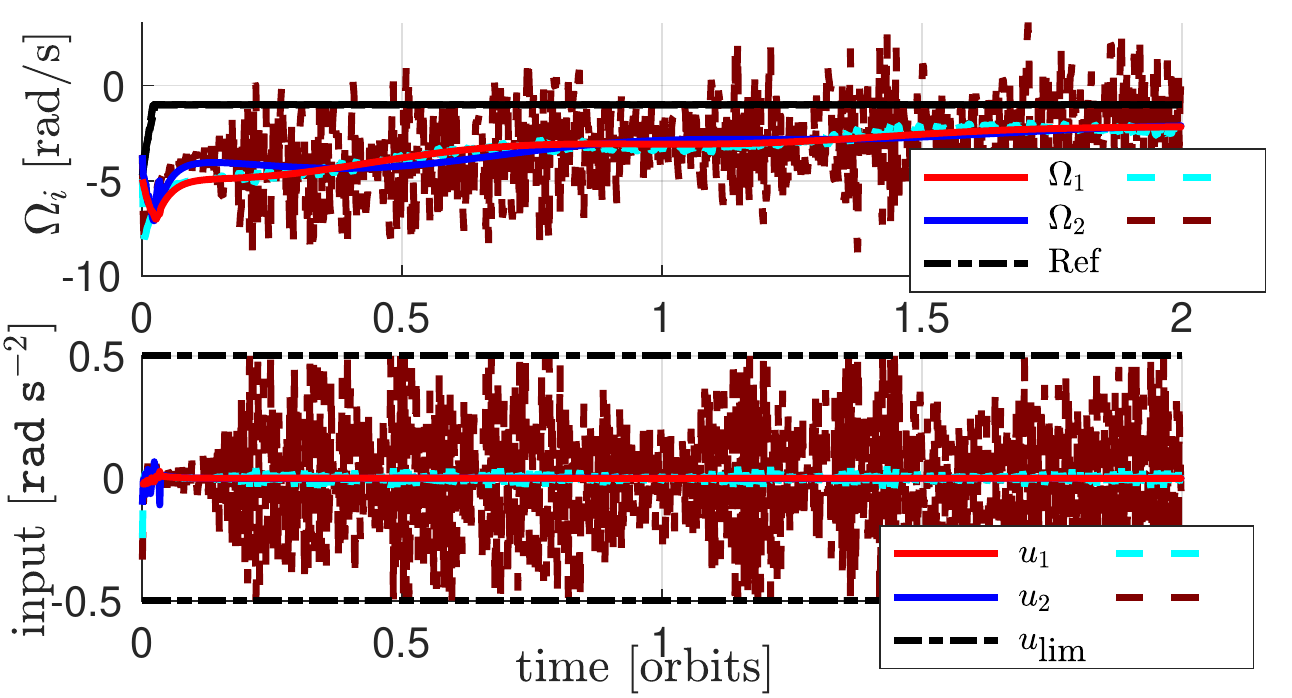}
\caption{State and input trajectories during a RW desaturation maneuver 
using TDMPC (dashed lines) or RG-TDMPC (solid lines) with adjusted terminal set to avoid oscillations. In both cases, $l\in[1,10]$ is selected from the uniform distribution.}
\label{fig:randDesat}
\end{figure}

As a more extreme scenario, suppose that at each time step a varying number of optimizer iterations is performed with $l$ chosen randomly from $\{1,2,\dots,10\}$ according to the uniform distribution so that half of the time $l\leq l_{min}$. This degrades the TDMPC performance as can be observed from the high amplitude oscillations in the control input and states in Figure \ref{fig:randDesat}. For RG-TDMPC, oscillations are avoided with a slight additional modification. Note that prior to $v_k = r$, updates leading to oscillations  are naturally rejected as large changes in control inputs would lead to pointing constraint violation. When $v_k =r$, pointing constraints are typically inactive but oscillations can be avoided by rejecting updates that result in $\bar x^{\tt ext}_{N_{\tt MPC}}$ far from the final set-point. This is achieved by checking (\ref{eq:IRG_refCondSub2}) with $v^+=r$, $N_{\tt RG} = N_{\tt MPC}$ and $c_F = 10^5$ in (\ref{eq:TSetSim}). Figure \ref{fig:randDesat} demonstrates the effectiveness of the modification.
\subsection{Zero-speed crossing avoidance}
\label{sec:compcMPC}
As stated in the introduction, during RW desaturation avoiding zero-speed crossing of the RWs can be desirable. We enforce this property by requiring that $$sign(x_{0,i}) x_{k,i} \geq 0.3,\quad i=7,\ldots,10,\; k\in \mathbb N.$$

\begin{figure}[ht]
\centering
\includegraphics[trim={0 0 0 0}, clip,scale=0.7]{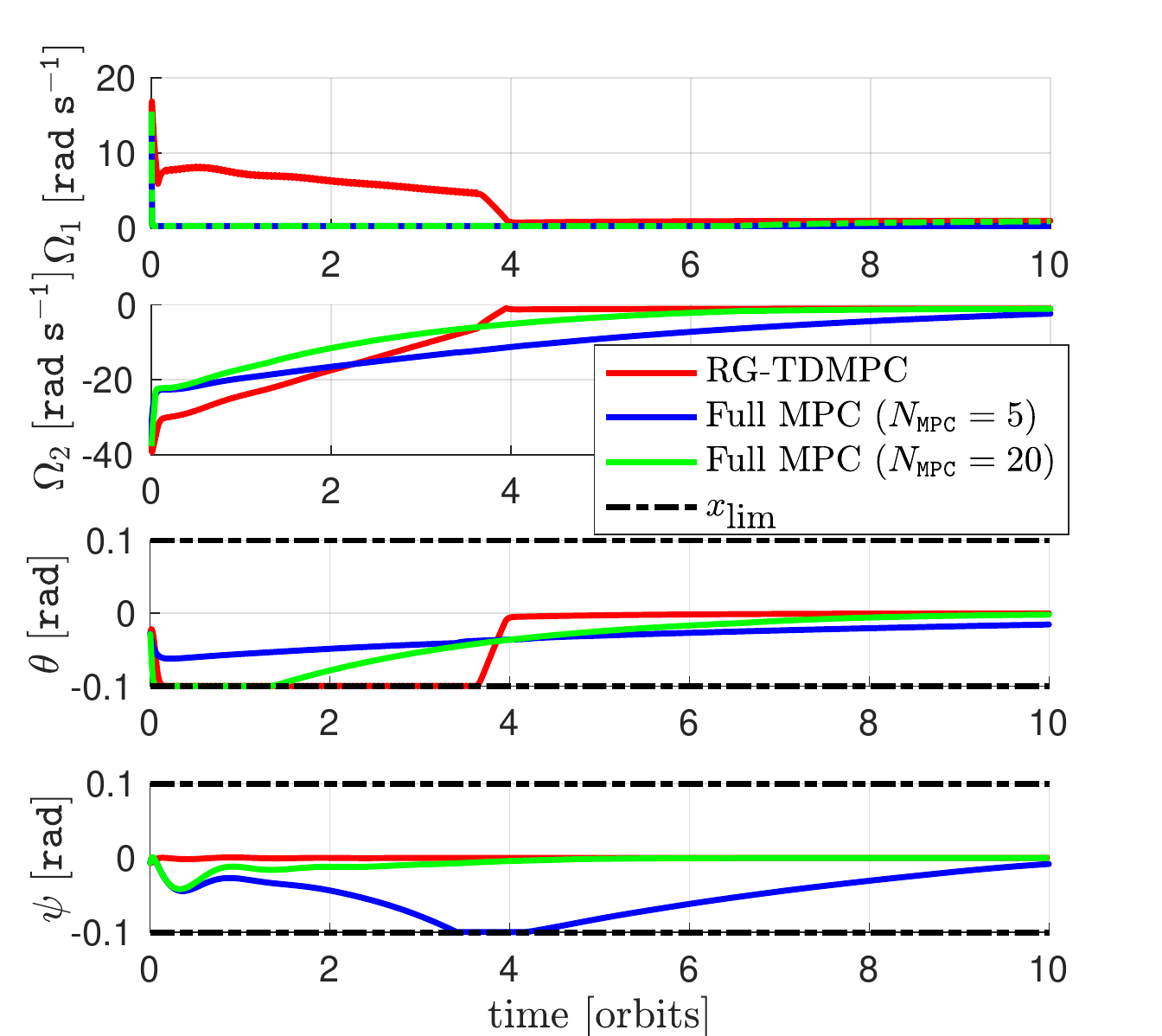}
\caption{State trajectories during RW desaturation maneuver from $x_0^\top = [-0.006,0.009,-0.023,0,-n,0,15.5,-37.7,15.1,38.1]$ to  $x^{eq}([-1\;1]^\top)$ using three controllers.
All state and input constraints are considered. Evolution of $\Omega_3,\;\Omega_4$ are similar to those of $\Omega_1,\;\Omega_2$, respectively. Inputs are well below the limits for most of the maneuver.}
\label{fig:4RWDesat_cstr}
\end{figure}

Figure \ref{fig:4RWDesat_cstr} shows the time histories of several states during the desaturation maneuver for RG-TDMPC ($l= l_{min}$ at all times) and for two input and state constrained MPCs, referred to as Full MPCs. The Full MPCs have horizons $N_{\tt MPC}=5$ and $N_{\tt MPC}=20$, respectively, and no terminal constraints. They use a damped generalized Newton's method developed by \cite{liao2018regularized}. 
One can observe that RG-TDMPC takes four orbits to end the maneuver, the Full MPC with $N_{\tt MPC} = 20$ takes six orbits and the Full MPC with $N_{\tt MPC} = 5$ takes ten orbits. The latter follows a very different trajectory from the other controllers: $\psi$ reaches values of 0.1 $\tt rad$ while $\theta$ remains smaller.
\subsection{Average input computation times}
We compare the computational load for four controllers: TDMPC, RG-TDMPC and the two Full MPCs presented above. For the Full MPCs and RG-TDMPC all constraints from Section \ref{sec:numVals} are considered.
To estimate computational load during a desaturation maneuver we compute the average input command computation time, $\bar t_{c}$ from input command computation times collected at each time step. We collect $\bar t_c$ for a set of 100 randomly generated initial conditions defined by attitude angles smaller than $0.05\;\tt rad$, angular velocities starting at the equilibrium and $x_7 = c_1+\delta_1,\;x_8 = c_2+\delta_2,\;x_9 = c_1+\delta_3,\;x_{10} = c_2+\delta_4$, where $|c_1|,\;|c_2| \leq 90$ and $|\delta_{i}|\leq 2.5$, $i=1,2,3,4$. Simulations were performed using Matlab on a laptop with a 2.3 GHz 8-Core Intel Core i9. 
Table \ref{fig:statTcompAv} shows mean and maximum value of the average computation time required to generate the inputs. RG-TDMPC, while slower than TDMPC, is one order of magnitude faster than the long horizon Full MPC and slightly faster than the short horizon Full MPC. Although not depicted, the short horizon Full MPC was prone to oscillations of inputs and states (in about 40\% of the simulations)  and, overall, required  longer maneuver times than the two other controllers. Note further that RG-TDMPC speed-up may be possible using tailored
methods for input constrained MPC such as in \cite{kogel2011fast} and
by using alternative terminal sets fitted to the constraints geometry.
\begin{table}[ht]
\resizebox{.5\textwidth}{!}{
\begin{tabular}{@{}ccccc@{}}
\toprule
 & TDMPC & RG-TDMPC & Full MPC ($N_{\tt MPC}=5$)  & Full MPC ($N_{\tt MPC}=20$) \\
$\bar t_{c}\;[{\tt ms}]$ (mean)& 0.41 & 2.34 & 2.95 & 15.70   \\
$\bar t_{c}\;[{\tt ms}]$ (max) & 0.65  & 3.91 & 6.56 & 35.25 \\\bottomrule
\end{tabular}%
}
\caption{Statistical values of $\bar t_{c}$ for different controllers (100 ICs considered).}
\label{fig:statTcompAv}
\end{table}

\section{Conclusion}
The paper demonstrated linear controllability of coupled spacecraft attitude and reaction wheel rotational dynamics for a spacecraft affected by the gravity gradient torques and equipped with four reaction wheels in pyramidal configuration (except for some singular configurations).  By exploiting controllability index, we have shown that in situations when the principal moments of inertia of the spacecraft satisfy $J_1\approx J_3$ or $J_2 \approx J_3$ the controllability is degraded; this suggests that an axisymmetric spacecraft should be aligned with its minor axis of inertia along the local vertical to facilitate RW desaturation. MPC strategies for RW desaturation subject to spacecraft pointing constraints have been considered. One of the proposed strategies, RG-TDMPC, which is based on an augmentation of Time Distributed MPC (TDMPC) with Incremental Reference Governor (RG) demonstrated the capability to satisfy pointing and zero speed crossing constraints at a fraction of the computational cost (estimated by the input computation time) of state and input constrained exact MPC with similar performance.

\section{Acknowledgments}
We thank Dominic Liao-McPherson for providing the implementation of the solver used in Section \ref{sec:compcMPC}


\begin{thebibliography}{10}

\bibitem{bryson1993control}
Arthur~Earl Bryson.
\newblock {\em Control of spacecraft and aircraft}, volume~41.
\newblock Princeton university press Princeton, New Jersey, 1993.

\bibitem{ACC23}
Miguel Castroviejo, Jordan Leung, and Ilya Kolmanovsky.
\newblock Reference governor for input-constrained {MPC} to enforce state
  constraints at lower computational cost, 2022.

\bibitem{CCTA22}
Miguel Castroviejo, Jordan Leung, and Ilya Kolmanovsky.
\newblock Suboptimal {MPC}-based spacecraft attitude control with reaction
  wheel desaturation.
\newblock In {\em 2022 IEEE Conference on Control Technology and Applications
  (CCTA), to be published}. IEEE, 2022.

\bibitem{diehl2005nominal}
Moritz Diehl, Rolf Findeisen, Frank Allg{\"o}wer, Hans~Georg Bock, and
  Johannes~P Schl{\"o}der.
\newblock Nominal stability of real-time iteration scheme for nonlinear model
  predictive control.
\newblock {\em IEE Proceedings-Control Theory and Applications},
  152(3):296--308, 2005.

\bibitem{garone2017reference}
Emanuele Garone, Stefano Di~Cairano, and Ilya Kolmanovsky.
\newblock Reference and command governors for systems with constraints: A
  survey on theory and applications.
\newblock {\em Automatica}, 75:306--328, 2017.

\bibitem{7330731}
Alberto Guiggiani, Ilya Kolmanovsky, Panagiotis Patrinos, and Alberto Bemporad.
\newblock Constrained model predictive control of spacecraft attitude with
  reaction wheels desaturation.
\newblock In {\em 2015 European Control Conference (ECC)}, pages 1382--1387,
  2015.

\bibitem{kogel2011fast}
Markus K{\"o}gel and Rolf Findeisen.
\newblock A fast gradient method for embedded linear predictive control.
\newblock {\em IFAC Proceedings Volumes}, 44(1):1362--1367, 2011.

\bibitem{kron2014four}
Aymeric Kron, Am{\'e}lie St-Amour, and Jean de~Lafontaine.
\newblock Four reaction wheels management: algorithms trade-off and tuning
  drivers for the proba-3 mission.
\newblock {\em IFAC Proceedings Volumes}, 47(3):9685--9690, 2014.

\bibitem{leve2015spacecraft}
Frederick~A Leve, Brian~J Hamilton, and Mason~A Peck.
\newblock {\em Spacecraft momentum control systems}, volume 1010.
\newblock Springer, 2015.

\bibitem{liao2018regularized}
Dominic Liao-McPherson, Mike Huang, and Ilya Kolmanovsky.
\newblock A regularized and smoothed fischer--burmeister method for quadratic
  programming with applications to model predictive control.
\newblock {\em IEEE Transactions on Automatic Control}, 64(7):2937--2944, 2018.

\bibitem{liao2020time}
Dominic Liao-McPherson, Marco~M Nicotra, and Ilya Kolmanovsky.
\newblock Time-distributed optimization for real-time model predictive control:
  Stability, robustness, and constraint satisfaction.
\newblock {\em Automatica}, 117:108973, 2020.

\bibitem{rawlings2017model}
James~Blake Rawlings, David~Q Mayne, and Moritz Diehl.
\newblock {\em {M}odel {P}redictive {C}ontrol: {T}heory, {C}omputation, and
  {D}esign}, volume~2.
\newblock Nob Hill Publishing Madison, WI, 2017.

\bibitem{wie1998space}
Bong Wie.
\newblock {\em Space {V}ehicle {D}ynamics and {C}ontrol}.
\newblock Aiaa, 1998.

\end{thebibliography}

\end{document}